# Polymer-derived SiOC Replica of Material Extrusion-based 3-D Printed Plastics

Apoorv Kulkarni[1,2], Gian Domenico Sorarù[1], Joshua M. Pearce[2,3,4, *]

1. Department of Industrial Engineering, University of Trento, Trento, Italy.

2. Department of Materials Science & Engineering, Michigan Technological University, MI, USA.

3. Department of Electrical & Computer Engineering, Michigan Technological University, MI, USA.

4. School of Electrical Engineering, Aalto University, Espoo, Finland

* corresponding author: pearce@mtu.edu

## Abstract

A promising method for obtaining ceramic components with additive manufacturing (AM) is to use a two-step process of first printing the artifact in polymer and then converting it to ceramic using pyrolysis to form polymer derived ceramics (PDCs). AM of ceramic components using PDCs has been demonstrated with a number of high-cost techniques, but data is lacking for fused filament fabrication (FFF)-based 3-D printing. This study investigates the potential to use the lower-cost, more widespread and accessible FFF-based 3-D printing of PDCs. Low-cost FFF machines have a resolution limit set by the nozzle width, which is inferior to the resolutions obtained with expensive SLA or SLS AM systems. To match the performance a partial PDC conversion is used here, where only the outer surface of the printed polymer frame is converted to ceramic. Here the FFF-based 3-D printed sample is coated with a preceramic polymer and then it is converted into the corresponding PDC sample with a high temperature pyrolysis process. A screening experiment is performed on commercial filaments to obtain ceramic 3-D prints by surface coating both hard thermoplastics: poly lactic acid (PLA), polycarbonate (PC), nylon alloys, polypropylene (PP), polyethylene terephthalate glycol (PETG), polyethylene terephthalate (PET), and co-polyesters; and flexible materials including: flexible PLA, thermoplastic elastomer and thermoplastic polyurethane filaments. Mass and volume changes were quantified for the soaking and pyrolysis steps to form a hollow ceramic skin. All 3-D printing materials extruded at 250 microns successfully produced ceramics skins of less than 100 microns. Details on the advantages and disadvantages of the different 3-D printing polymer precursors are discussed for this processing regime. The results are analyzed and discussed in order to provide guidance for more widespread application of AM of PDCs.

**Graphical Abstract**

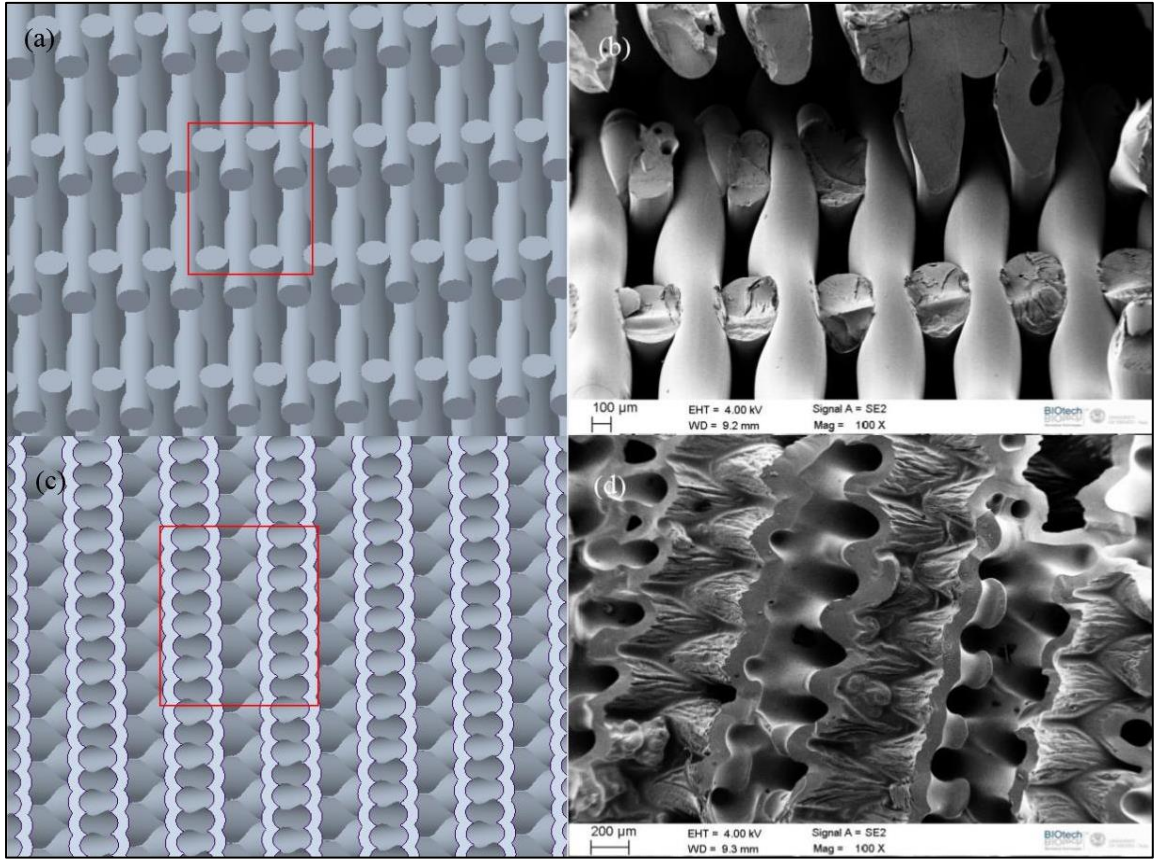

**Highlights**

- Additive manufacturing of polymer derived ceramics with fused filament fabrication
- Producing ceramics with hollow struts by surface coating with preceramic polymers
- Creating a multi-level porous system with stable geometry
- All 3-D printing materials produced ceramics skins of less than 100 microns

## 1. Introduction

Ceramic additive manufacturing (AM) is now well established and has been demonstrated with several technologies including: selective laser sintering (SLS) [1,2,3], stereolithography (SLA) [4], laminated object manufacturing (LOM) [5,6], direct ink writing (DIW) [7], binder jetting [8], directed energy deposition (DED) [9] and material (paste) extrusion [10,11].

Each of these methods has inherent challenges. SLS using powder substrates, results in low-density parts, and uses high energy densities requiring sophisticated laser equipment, which is costly [12]. Even though ceramic SLA provides high resolution, the cost of the feedstock (i.e. ceramic particles mixed with photocurable polymer) is high [13]. In LOM, decubing is the most difficult task as the ceramic parts can be severely damaged during this step as cutting occurs though strong and weak interfaces of ceramic and the adhesive [14]. The parts also need finishing after decubing that adds to the cost of the process [14]. Optimizing the rheology and in turn the layer uniformity is challenging in DIW [15]. Binder jetting is a multistep process where the binder needs to be removed from the manufactured part and it needs to be post processed to increase the density. The post-processing step can include sintering or isostatic pressing, which increases the total processing cost and time as well as the capital cost of the equipment [16]. DED of ceramics is difficult as it requires the feedstock to be heated to form a molten pool to fuse into the previous layer, which is a challenging physical process for many ceramics. In addition, DED requires a slow deposition rate increasing the build time of the process [17]. Finally, even though ceramic paste extrusion is inexpensive, it limits the resolution of the extrusion and the ability to manufacture complex geometric shapes due to rheological properties of the paste [18]. In addition, paste extrusion is hampered by drying of the paste inside the nozzle obstructing the extrusion and curling of the extruding paste [19]. Most detrimentally, cracks can also be introduced while drying the deposited paste compromising the mechanical properties of the part [20].

One approach that can enable AM of ceramics is to use the polymer pyrolysis route also known as polymer derived ceramic (PDC) route. It uses a two-step process of first printing the artifact using a polymer (preferably a preceramic polymer) AM (a much more mature AM material system that is now widespread among even consumers [21-23]) then converting the 3-D printed structure into the polymer derived ceramic object via high temperature pyrolysis in controlled atmosphere. The PDC approach has many advantages compared to the conventional ceramic powder processing methods such as the lower processing temperatures, higher resolution and the possibilities to process multicomponent and multifunctional ceramic systems [24-27].

AM using PDCs has been demonstrated with powder based SLS, SLA and DIW [28,29]. DIW was effective using a solution of isopropyl alcohol and silicone resin powder [30]. A similar study was conducted to manufacture SiOC ceramics with ordered porosity [31]. SiOC ceramic microcomponents have also been manufactured with SLA by using photosensitive preceramic polymers and LED curing technology [32]. PDC parts were also manufactured by laser curing of SiC mixed with polysiloxane [33]. The primary challenges with 3-D printing ceramics using DIW is controlling the rheology of the ink mixture and stability of the deposited part, while with laser curing and SLA are inhibited from widespread adoption by the required expensive equipment. However, the most widespread form of AM is fused filament fabrication (FFF)-based 3-D printing made possible by the open source release of the self-replicating rapid prototyper (RepRap) project [34-36]. There is a dearth of knowledge on the potential for FFF-based 3-D printing to fabricate ceramic components using the PDC method.

To fill this knowledge gap and overcome the above-mentioned challenges with other methods, this study investigates the potential to use the lower-cost, more widespread and accessible FFF-based 3-D printing of PDCs. Low-cost FFF machines have a resolution limit set by the nozzle width, which is much larger than the resolutions obtained with expensive SLA of SLS AM systems. To match the performance of more expensive and less accessible machines an FFF 3-D printed structure is replicated by coating and pyrolysis where the 3-D printed structure is coated with preceramic polymer. During pyrolysis, the organic 3-D printed polymer is decomposed, and preceramic polymer is converted to ceramic resulting in a high-resolution structure. A screening experiment is conducted based on earlier work replicating polyurethane foams [37] and 3-D printed lattice with acrylic resin with SLA [38]. The PDC replica of a 3-D printed polymeric structure can, in principle, be possible if the pre-ceramic polymer swells and coats the struts of the organic polymeric structure. In this case, the resulting ceramic object will have dense struts as in the case of the ceramic foams obtained from the impregnation of polyurethane foams. If the pre-ceramic polymer mainly coats without swelling the struts of the 3-D printed object then the resulting porous ceramic object will have two interconnected porosity channels, one originating from the decomposition of the organic polymer and the other one due to the porosity between the original 3-D printed polymeric struts.

In the present work we report the results of screening experiments conducted using the following commercial filaments both from the hard (poly lactic acid (PLA), polycarbonate (PC), nylon alloys, polypropylene (PP), polyethylene terephthalate glycol (PETG), polyethylene terephthalate (PET), and co-polyesters ) and soft flexible (flexible PLA, thermoplastic elastomer and thermoplastic polyurethane filaments) thermoplastic materials.

In this experiment, samples were 3-D printed from filament materials, soaked in a preceramic polymer solution, dried at room temperature and then pyrolyzed to obtain a ceramic frame. Samples were characterized with scanning electron microscopy (SEM) to image and quantify the resultant 3-D hollow structures.

The results are analyzed and discussed in order to provide guidance for more widespread application of additive manufacturing of PDCs.

## 2. Materials and Methods

The following filaments material (with suppliers) were tested: PLA (Ultimaker), Polycarbonate (eSUN), Ally 910 (Taulman), PETG (eSUN), t-glase (Taulman), Inova 1800 (Chromastrand), nGen (Colorfabb), polypropylene (Verbatim), nGen Hard (Colorfabb) PRO series TPU (Matterhackers), Pro series TPE (Matterhackers), nGen Flex (Colorfabb), NinjaFlex (NinjaTek), PORO-LAY LAYFOMM 60, Rubberlay Solay, Flexfill 98A (Fillamentum), Flexsolid (Madesolid), PCTPE (Taulman), and SoftPLA (Matterhackers). The filaments were 3-D printed on a Lulzbot TAZ 6 (Aleph Objects) using an E3D SL 0.25mm extruder and standard 0.5 mm extruder using the parameters summarized in Table 1 for hard thermopolymers and Table 2 for flexible thermopolymers.

The samples were designed as 10mm × 10mm × 10mm cubes in OpenSCAD 2015.03. The cubes were then prepared as a mesh with line type infill and 40% infill density (distance between two lines is 625 microns) using Cura Lulzbot 3.6.3 open source slicer.

| Hard | | | | | | | | |
|---|---|---|---|---|---|---|---|---|
| Filament | PLA (Ultimaker) | Polycarbonate (eSUN) | Alloy 910 (Taulman) | PETG (eSUN) | t-glase PETT (Taulman) | Inova 1800 (Chroma strand) | Polypropylene (Verbatim) | nGen (Colorfabb) |
| Color | Clear | Black | Natural | Clear | Blue | Red | Clear | Green |
| Printing Temperature upper limit (°C) | 210 | 290 | 260 | 245 | 230 | 240 | 260 | 240 |
| Printing Temperature lower limit (°C) | 180 | 250 | 250 | 220 | 210 | 230 | 230 | 220 |
| Bed Temperature (°C) | 55 | 120 | 55 | 55 | 60 | 50 | print on polypropylene tape | 85 |
| Printing Speed (mm/sec) | 40 | 30 | 45 | 17 | 22 | 35 | 15 | 55 |
| Bottom layer print speed (mm/sec) | 15 | 15 | 15 | 15 | 10 | 15 | 15 | 15 |
| Layer height (mm) | 0.2 | 0.25 | 0.2 | 0.2 | 0.2 | 0.2 | 0.25 | 0.2 |
| Bottom layer height (mm) | 0.4 | 0.4 | 0.4 | 0.4 | 0.4 | 0.4 | 0.4 | 0.4 |
| Nozzle diameter (mm) | 0.25 | 0.5 | 0.25 | 0.25 | 0.25 | 0.25 | 0.5 | 0.25 |

Table 1: List of hard thermoplastic filaments and their printing parameters

| Flexible | | | | | | | | | | |
|---|---|---|---|---|---|---|---|---|---|---|
| Filament | PRO Series TPU (Matterhackers) | PRO Series TPE (MatterHackers) | nGen Flex (ColorFabb) | NinjaFlex (Ninjatek) | PORO-LAY LAY-FOMM 60 Porous Filament | RUBBERLAY SOLAY Elastic Filament | Flexfill 98A (Fillamentum) | FlexSolid (Madesolid) | PCTPE (Taulman) | Soft PLA (Matterhackers) |
| Color | Blue | Red | Dark Grey | Green | While | White | Black | Black | Natural | White |
| Printing Temperature upper limit (°C) | 260 | 235 | 260 | 235 | 230 | 195 | 220 | 250 | 245 | 230 |
| Printing Temperature lower limit (°C) | 240 | 230 | 240 | 225 | 220 | 175 | 200 | 230 | 230 | |
| Bed Temperature (°C) | 60 | 50 | 80 | 45 | no heated bed | use glue stick | 50 | 70 | 50 | 45 |
| Printing Speed (mm/sec) | 15 | 15 | 15 | 15 | 15 | 5 | 15 | 15 | 15 | 15 |
| Bottom layer print speed (mm/sec) | 5 | 5 | 10 | 10 | 5 | 5 | 10 | 5 | 10 | 10 |
| Layer height (mm) | 0.2 | 0.2 | 0.2 | 0.2 | 0.2 | 0.2 | 0.2 | 0.2 | 0.2 | 0.2 |
| Bottom layer height (mm) | 0.25 | 0.25 | 0.25 | 0.25 | 0.25 | 0.25 | 0.25 | 0.25 | 0.25 | 0.25 |
| Nozzle diameter (mm) | 0.25 | 0.25 | 0.25 | 0.25 | 0.25 | 0.25 | 0.25 | 0.25 | 0.25 | 0.25 |

Table 2: List of soft flexible thermoplastic filaments and their printing parameters

A commercial liquid polysiloxane, precursor for silicon oxycarbide (SiOC) ceramics, (Polyramic SPR-036, Starfire Systems) was selected. Platinum divinyltetramethyldisiloxane complex, ~Pt 2% in xylene (CAS number: 68478-92-2, Sigma-Aldrich, St. Louis, MO, USA) was further diluted to 0.1% before using as a catalyst. Pt catalyst promotes the crosslinking of the preceramic polymer *via* hydrosilylation reaction between the Si-H and the C=C moieties present in the silicon polymer [39]. 100 microliters of the catalyst solution per 1 gram of preceramic polymer were used. The preceramic polymer was dissolved in acetone (50:50 by weight). Acetone was selected as a solvent as it is less detrimental to the various polymers used in this experiment compared to other solvents [40]. The 3-D printed structures were soaked for a 30 minutes in the SPR-036/acetone/Pt solution. The samples were then taken out and left to dry in air at room temperature for 24 hours on a ceramic foam before pyrolysis.

The impregnated samples were then pyrolyzed in an alumina tube furnace (Thermolyne 5400 High Temperature tube furnace) at 1200°C in nitrogen flow (400 cc/min) with a heating rate of 10°C/min with 1 hour dwelling at 1200°C. The samples were then cooled freely to the room temperature. To remove any oxygen and moisture from the system, the furnace was purged for 45 minutes before heating, with the samples inside the tube [41].

To characterize the polymer-to-ceramic transformation, the samples were weighed with a digital scale (Mettler AT20, ± 0.00006 gm), and their volume was measured with a digital Vernier caliper (VINCA, ± 0.01 cm) before and after pyrolysis. The morphology of 3-D printed polymeric samples and of the corresponding ceramic samples after pyrolysis characterized by scanning electron microscopy (Zeiss Supra 40) after coating the samples with a thin Pt/Pd film. In order to get a clean fracture surface, the 3-D printed polymeric samples were dipped in liquid nitrogen before cutting them with a knife.

The decomposition process of selected (Ninjaflex, PRO TPU, polycarbonate, Alloy 910, PETG and Rubberlay) 3-D printed polymeric samples (not impregnated with the pre-ceramic polymer), and of the starting SPR-036 polymer before and after addition of the Pt catalyst (100 µl of Pt for 1 g of Si-polymer), was studied recording differential thermal analysis (DTA) and thermo-gravimetric analysis (TGA) using a Netzsch STA 409 equipment (Netzsch Gmbh, Selb, Germany) at 10°C/min up to 1,200°C in flowing $N_2$ (150 cc/min).

## 3. Results

A weight increase ranging from 6% to 35% percent for hard filaments and 8% to 30% for flexible filaments was observed after soaking and drying the samples as seen in Figure 1.

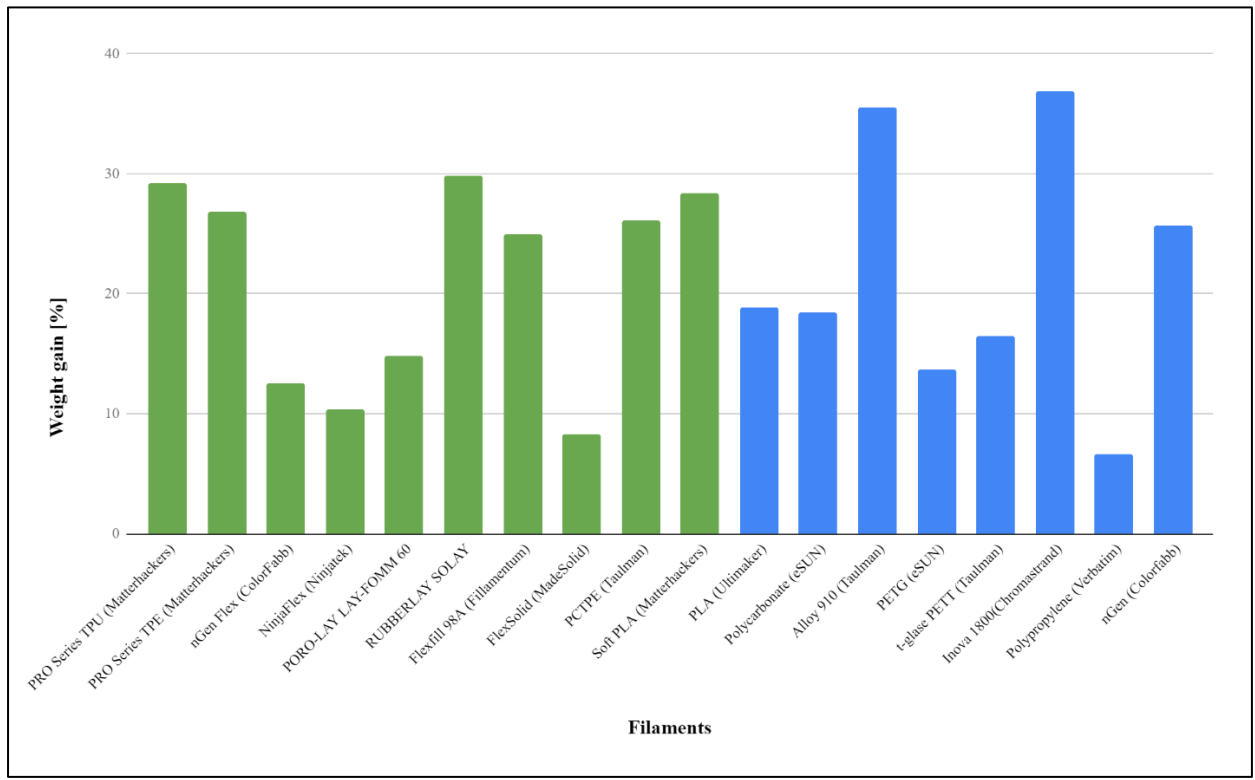

Figure 1: Percentage weight gain of 3-D printed samples after soaking in preceramic polymer solution and drying.

The materials that increased in weight the most were Nylon alloy Alloy 910 (Taulman), Soft PLA (Matterhackers) and thermoplastic elastomers PRO TPU, PRO TPE (Matterhackers), Rubberlay. While the materials that gained the least amount were PP (Verbatim) and PETG (eSUN). The weight increase is the result the Si-based compound forming a coating on the surface of the 3-D printed filament struts. No swelling of the samples was measured after drying therefore suggesting that the main reason for the observed weight increase of the samples after soaking in the pre-ceramic polymer solution was the formation of a coating on the surface of the 3-D printed struts.

TGA curves recorded on selected 3-D printed polymeric structures and their derivative (DTG) are shown in Figure 2.

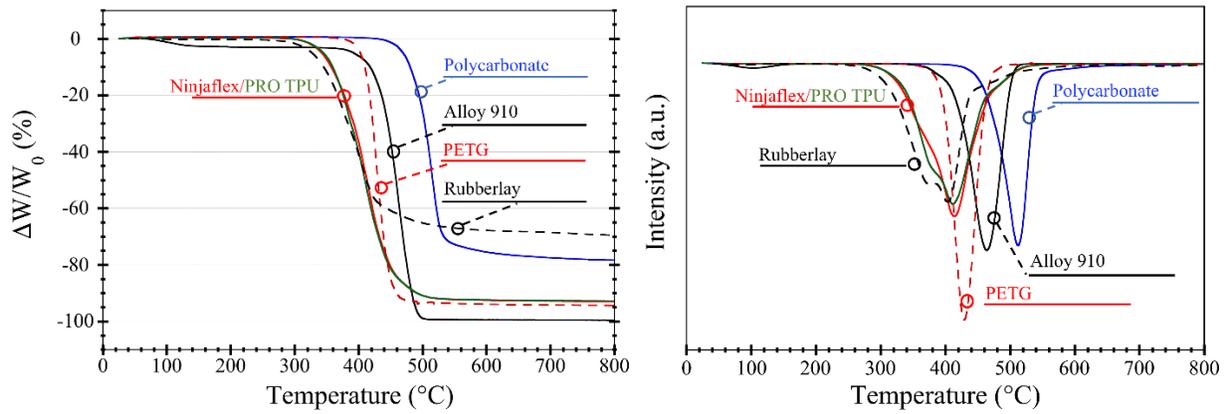

Figure 2. DTA (left) and TGA (right) curves recoded on selected 3-D printed polymeric samples.

The pyrolysis of the polymeric structures leads to an almost complete decomposition (weight loss above 95%) and only for two polymers, namely polycarbonate and Rubberlay, the carbonaceous residue is of the order of 20-30%. The temperature range in which this decomposition occurs goes from a minimum of ca 300°C to a maximum of ca 550°C. At these temperatures the pre-ceramic polymer is already crosslinked, and its pyrolysis leads to a volume shrinkage (and weight loss) without losing the initial shape. Indeed, DTA/TGA experiments performed on the as received SPR-036 and on the siloxane resin after addition of 100 mL/g of Pt catalyst (see Figure 3) indicate that crosslinking occurs readily at around 100 °C with a strong exothermic effect. After crosslinking, the weight loss of the pre-ceramic polymer is reduced to ca 30% compared to the 45% of weight loss associated to the non-crosslinked polymer. Not only the Pt-catalyzed siloxane polymer crosslinks before the decomposition of the 3-D printed structure but also before the melting of the polymeric struts (all the polymeric filaments used in the experiment have their melting region starting well below 275°C). Accordingly, without using catalyst, the polymeric samples would lose their structure and start melting before the preceramic is cured. In that case, a replicated structure of ceramic will not be obtained after pyrolysis.

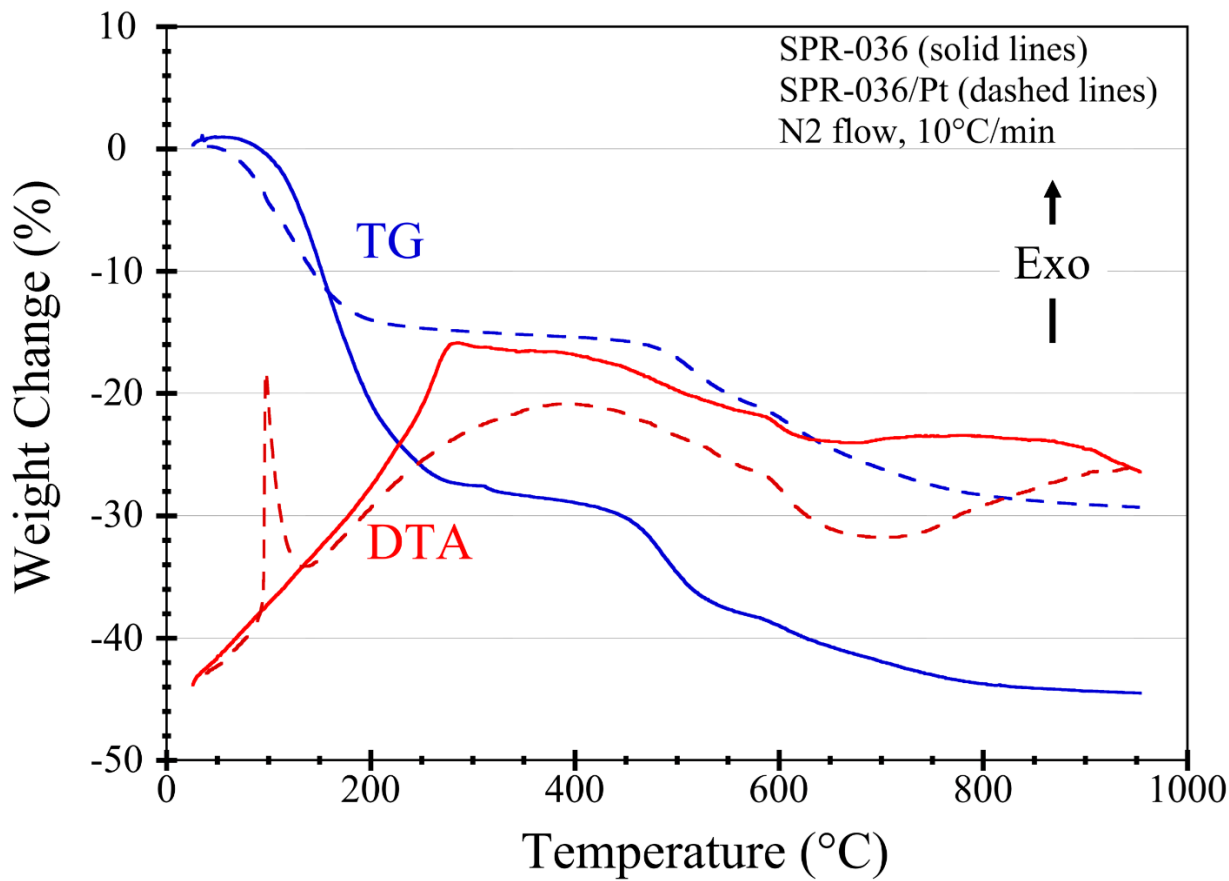

Figure 3. DTA/TGA curves recorded on the "as-received" SPR-036 and on the same polymer after addition of 100 mL/g of Pt catalyst (0.1%).

The volume shrinkage after pyrolysis measured for the impregnated 3-D samples (Figure 5) ranges from 33% to 67% and the corresponding weight loss (Figure 4) varies between 50% and 93%. The weight loss is the combined result of the thermal decomposition of the organic polymer and of the Si-based pre-ceramic compound. Since the decomposition of the organic polymer occurs with a nearly 95-100% weight loss the percentage of ceramic residue is directly related to the relative amount of the SPR-036 polymer coated on the surface of the 3-D printed polymeric objects. A high weight loss value during pyrolysis should be correlated to a thinner coating and should lead to thin ceramic structures. The sample that shows the lowest weight loss during pyrolysis (50%) is the Rubberlay, which must also be due to the high carbonaceous residue that this polymer shows (ca 30 wt%, see Figure 2).

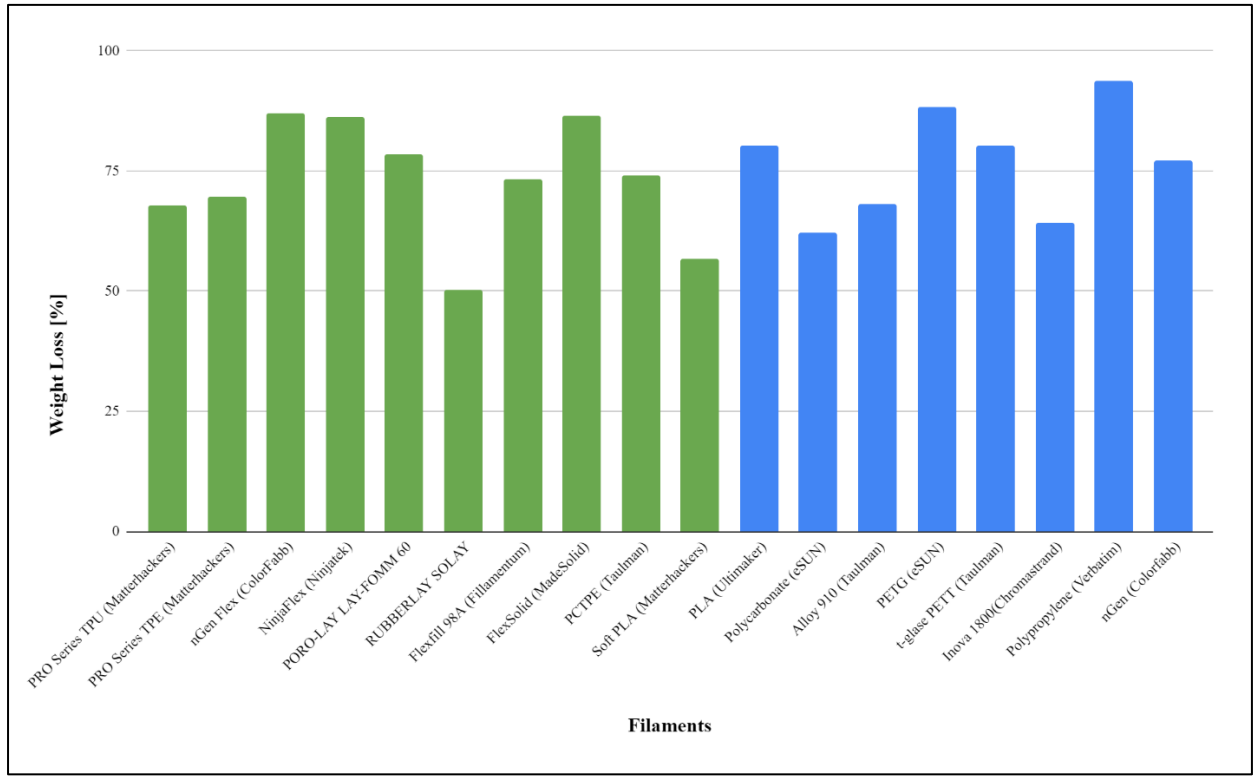

Figure 4: Percentage weight loss measured for the 3-D printed impregnated samples during the polymer-to-ceramic conversion.

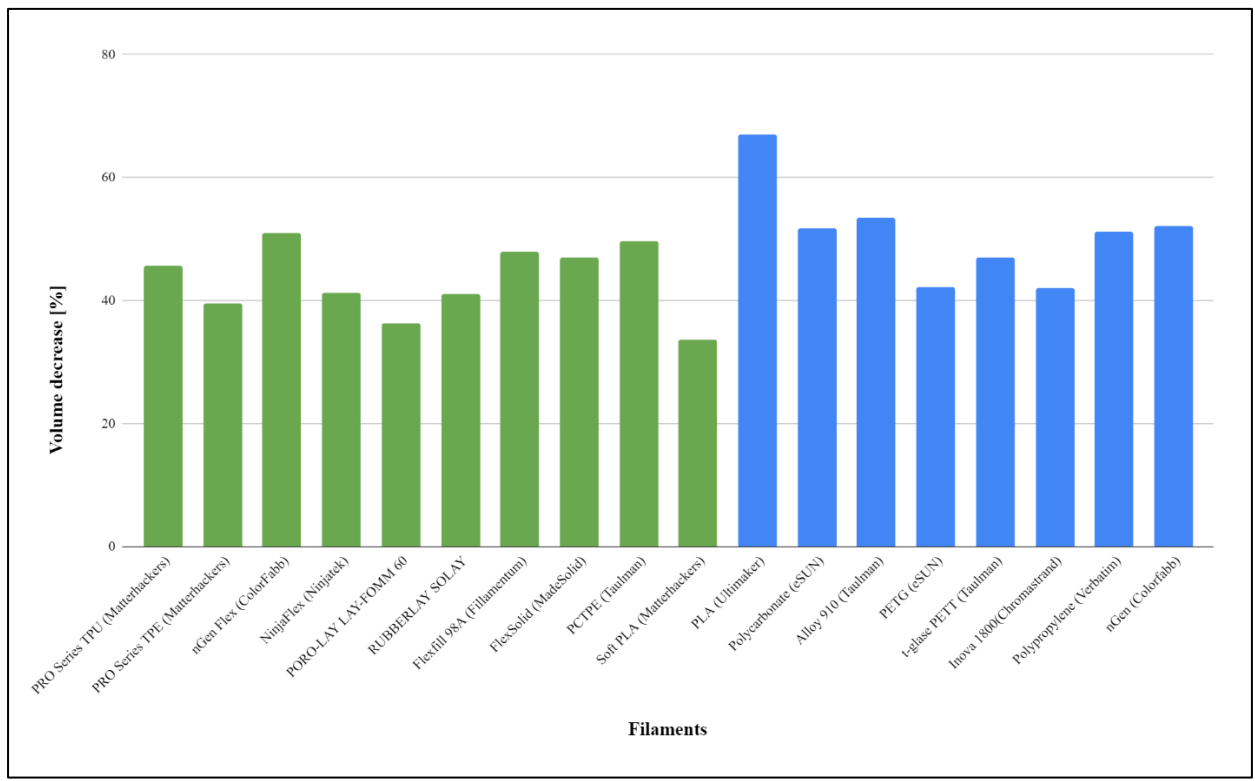

Figure 5: Percentage volume shrinkage measured for the 3-D printed impregnated samples during the polymer-to-ceramic conversion

The ceramic samples obtained after pyrolysis are compared to the initial polymeric ones in Figures 6-23. The ceramic samples held their structure without any major visible damage. However, depending on the type of polymer used in the FFF process there was some small distortion. PLA and PET (Figure 6 and10) displayed some highly reflective areas that can be ascribed to the excess pre-ceramic solution forming a continuous film on the face of the 3-D cube. In a similar manner, PC and PETG in Figure 7 and 9 shows some extra materials on the exterior that can be explained as extra preceramic polymer retained on the surface of the sample. This type of defect can be removed by further optimizing the soaking/drying process. Several materials provided nearly perfect replica of the structure such as Alloy 910 (Figure 8), PP (Figure 12), nGen (Figure 13), Pro TPU (Figure 14), Pro TPE (Figure 15), Ninjaflex (Figure 17), Poro-Lay LayFomm (Figure 18), Rubberlay Solay (Figure 19), FlexSolid (Figure 21), PCTPE (Figure 22), and Soft PLA (Figure 23). NGen Flex (Figure 16) and Flexfill 98A (Figure 20) had a non-uniform distortion of shape.

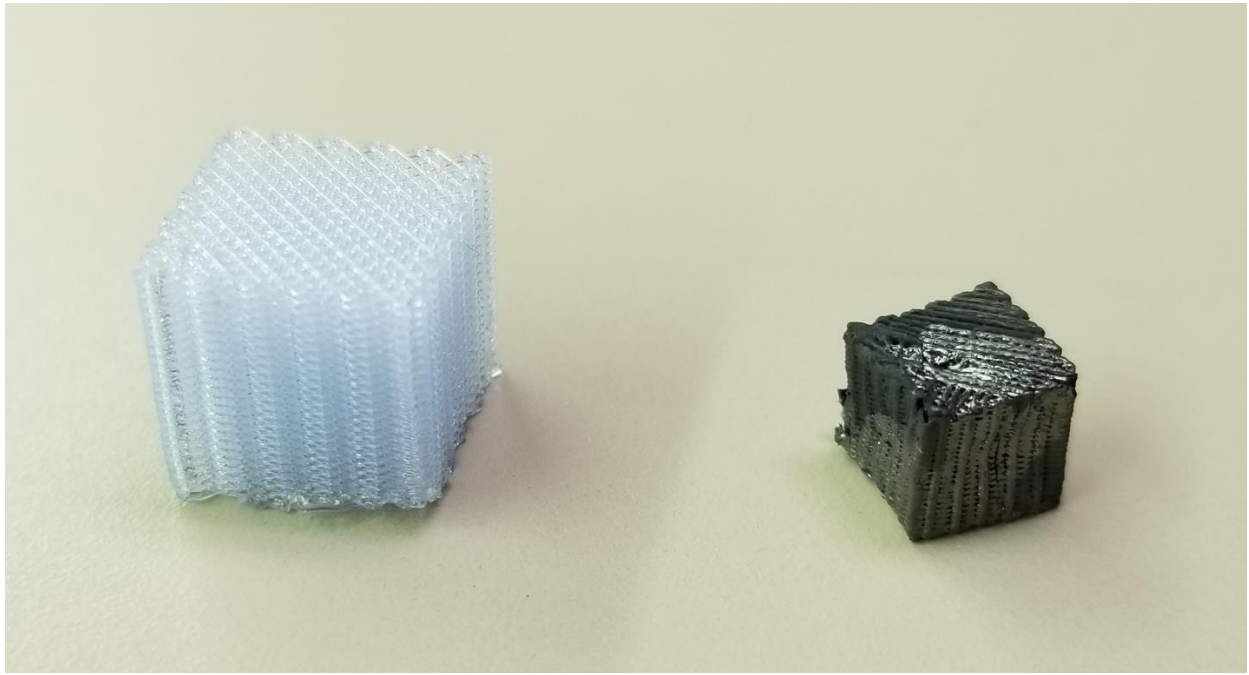

Figure 6: PLA (Ultimaker)

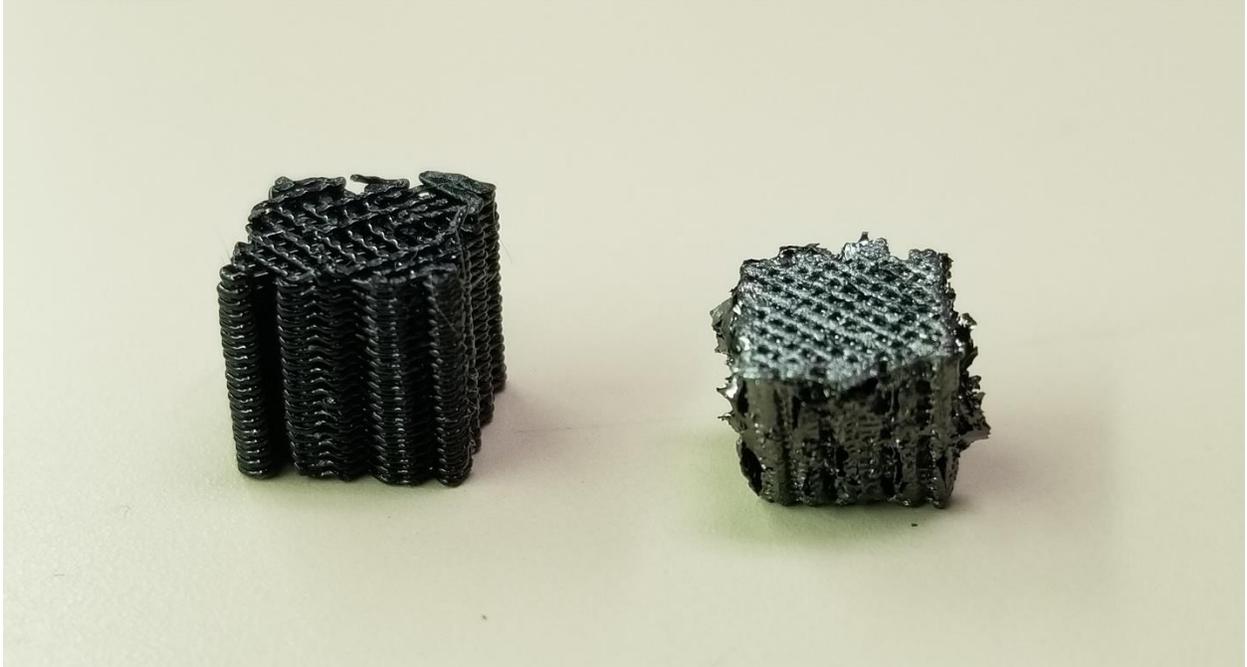

Figure 7: Polycarbonate (eSUN)

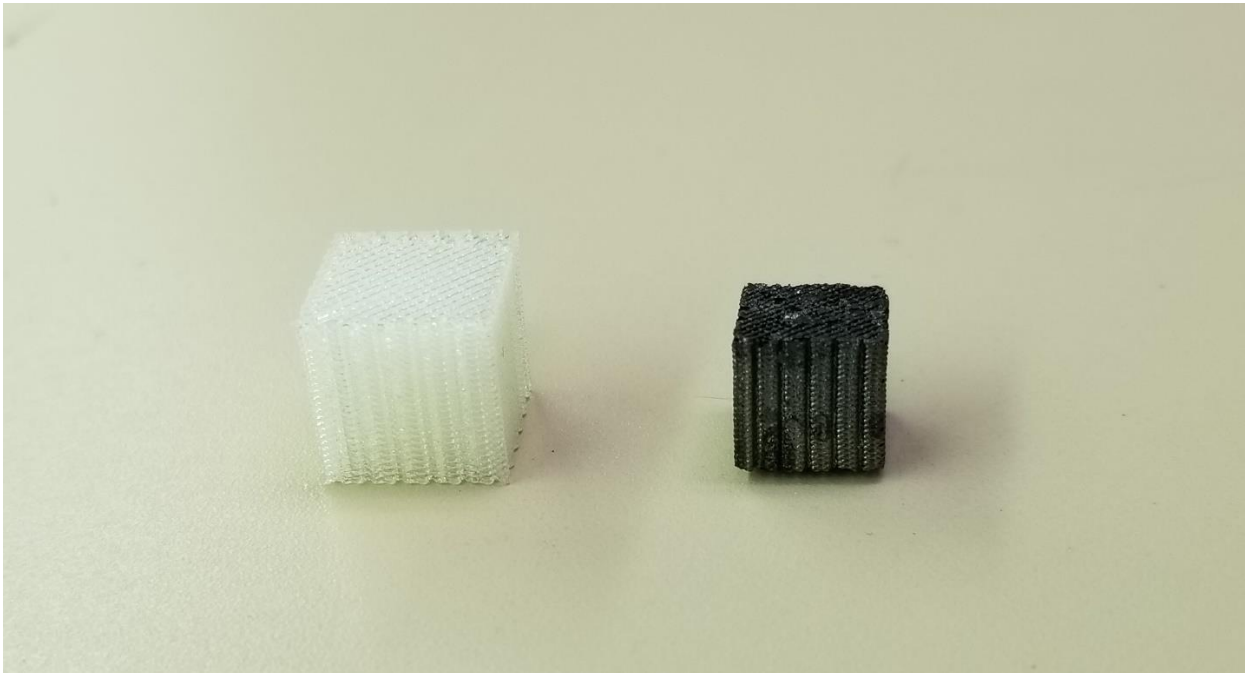

Figure 8: Alloy 910 (Taulman)

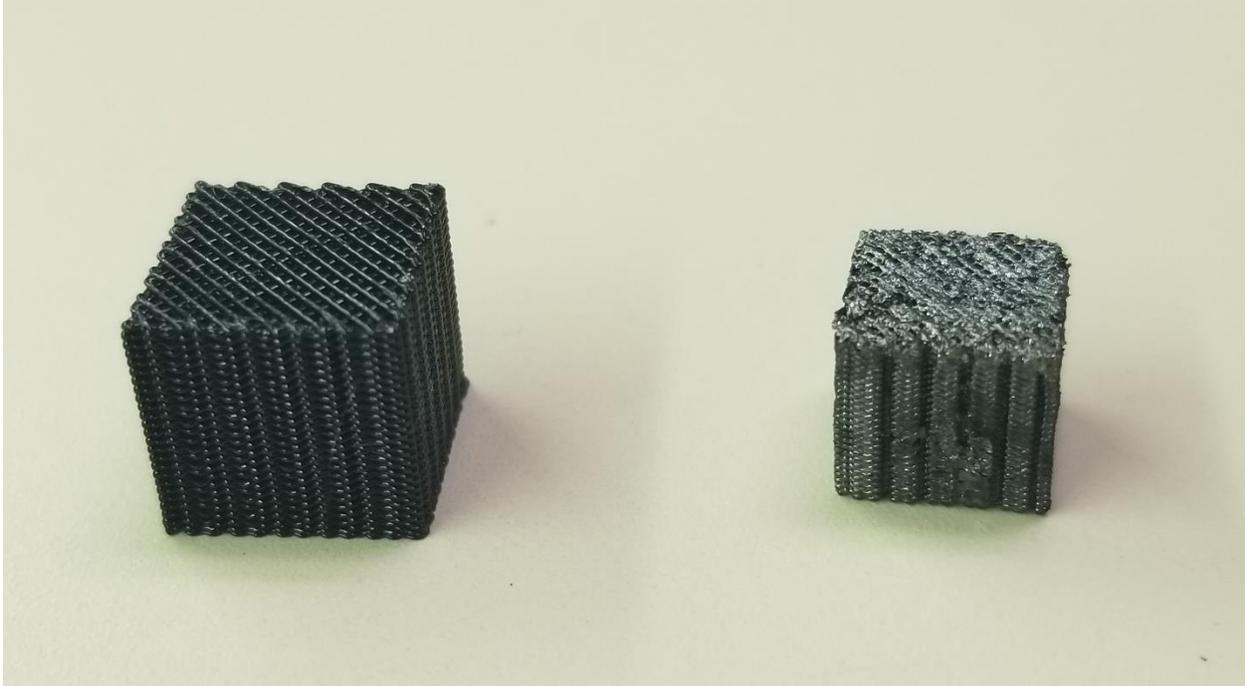

Figure 9: PETG (eSUN)

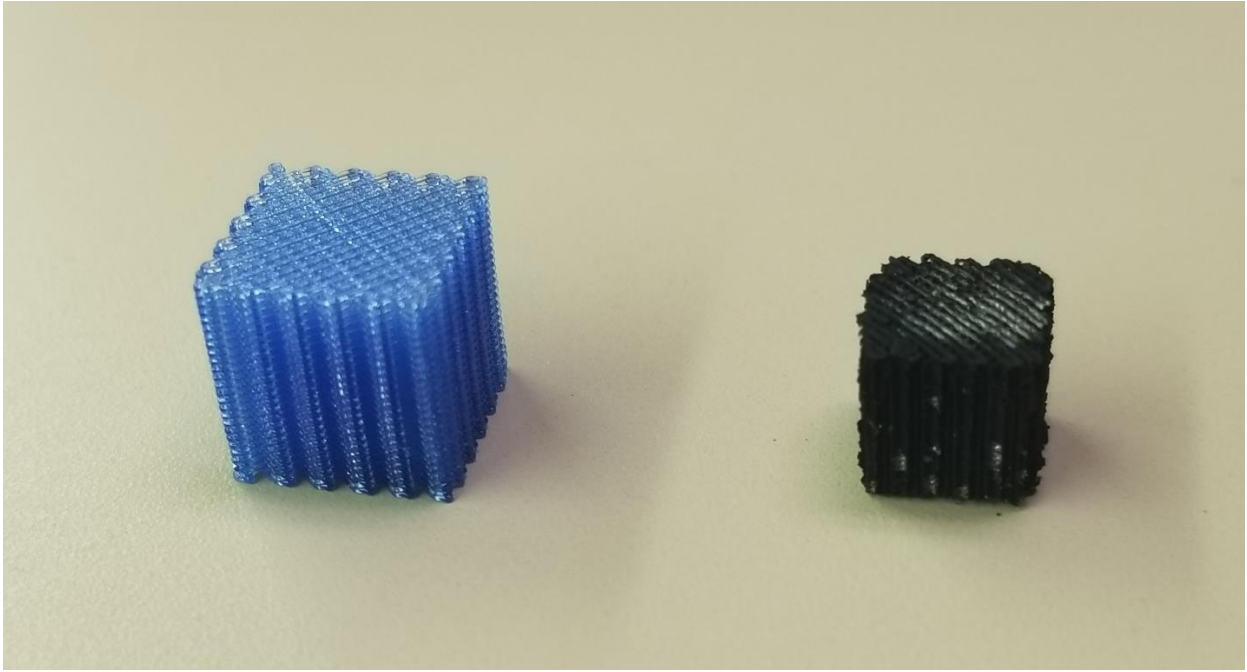

Figure 10: t-glase PETT (Taulman)

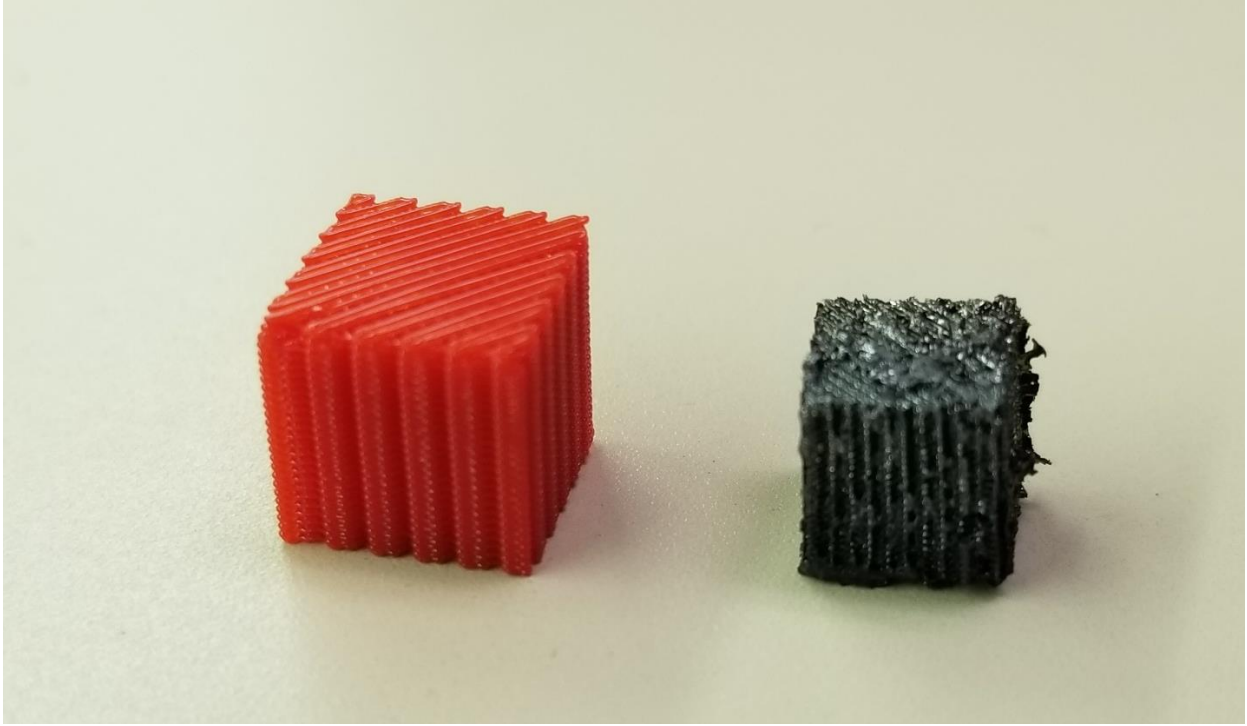

Figure 11: Inova 1800 (Chroma Strand)

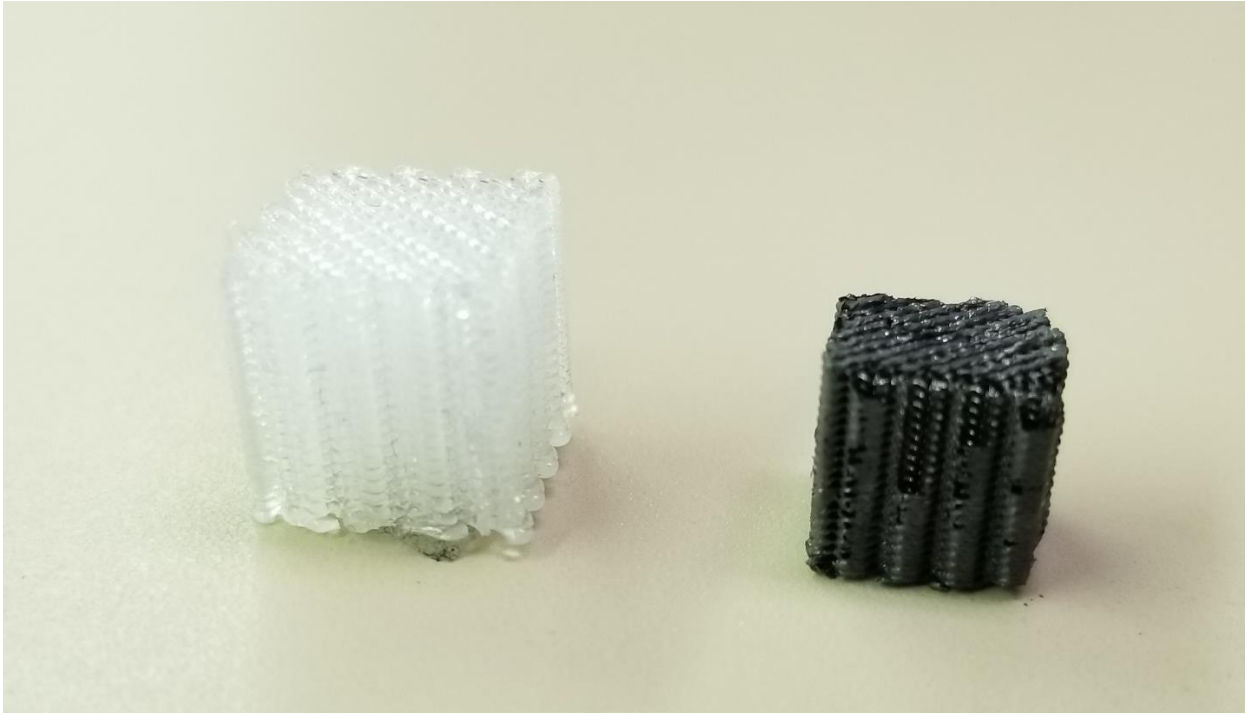

Figure 12: Polypropylene (Verbatim)

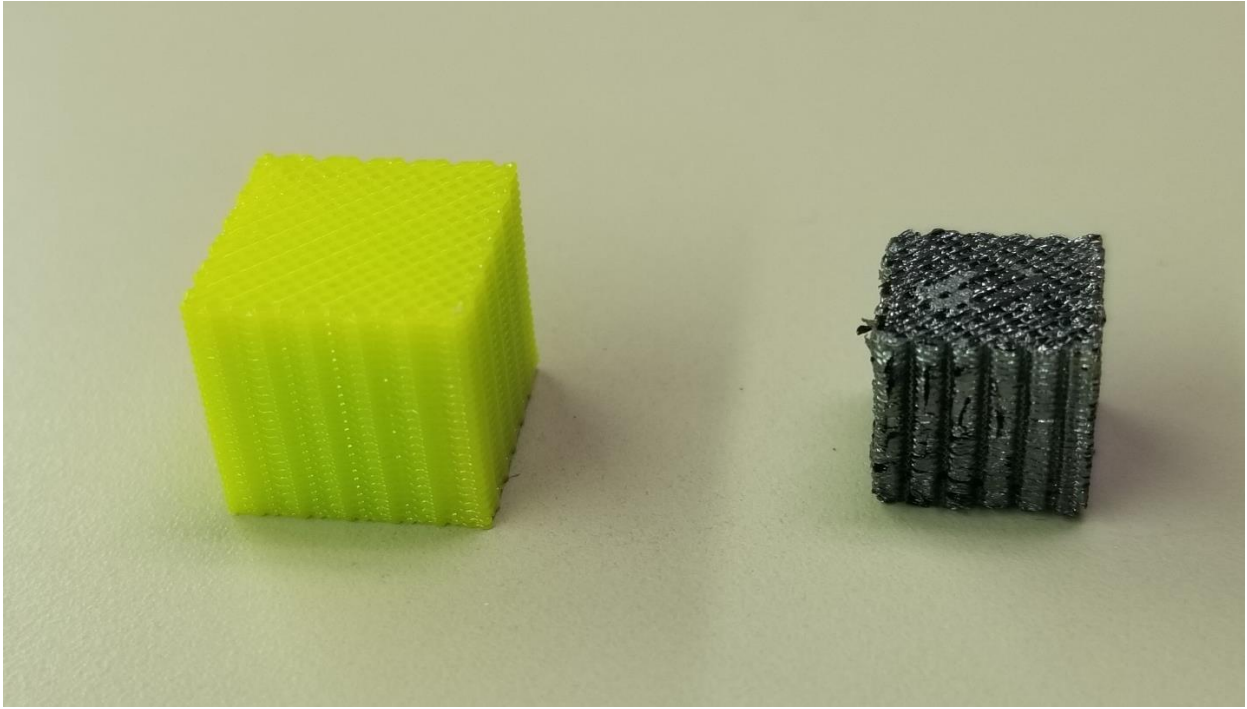

Figure 13: nGen Hard (Colorfabb)

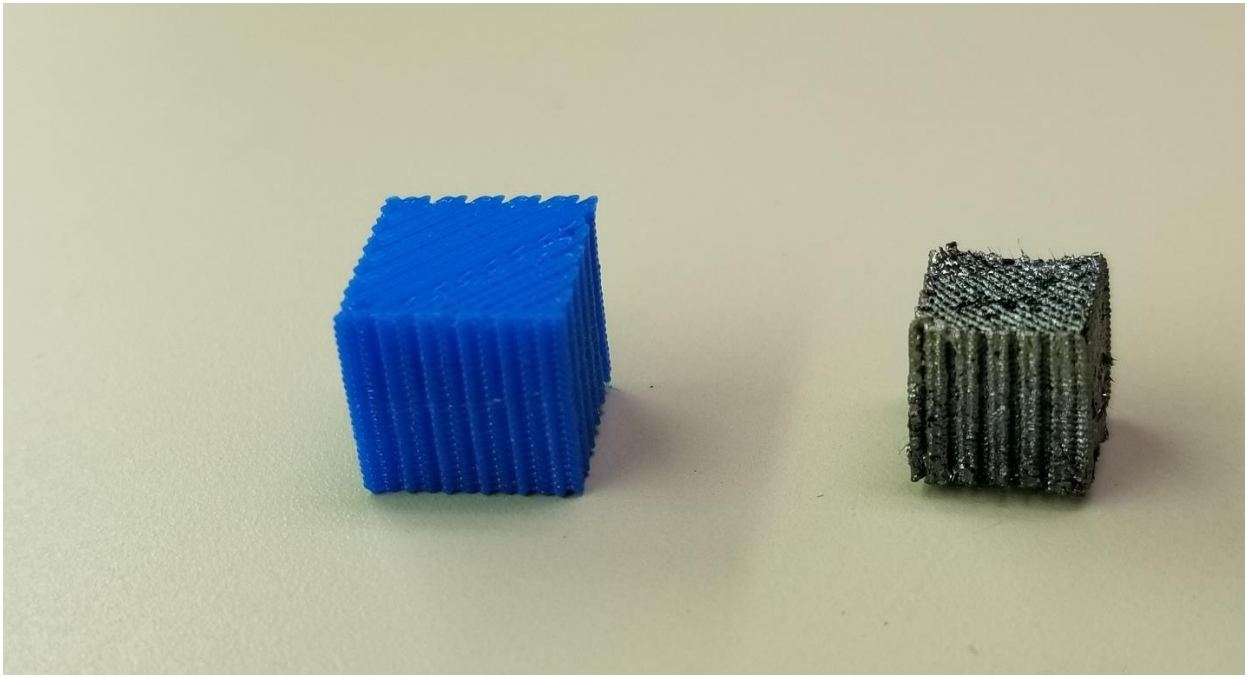

Figure 14: PRO TPU (Matterhackers)

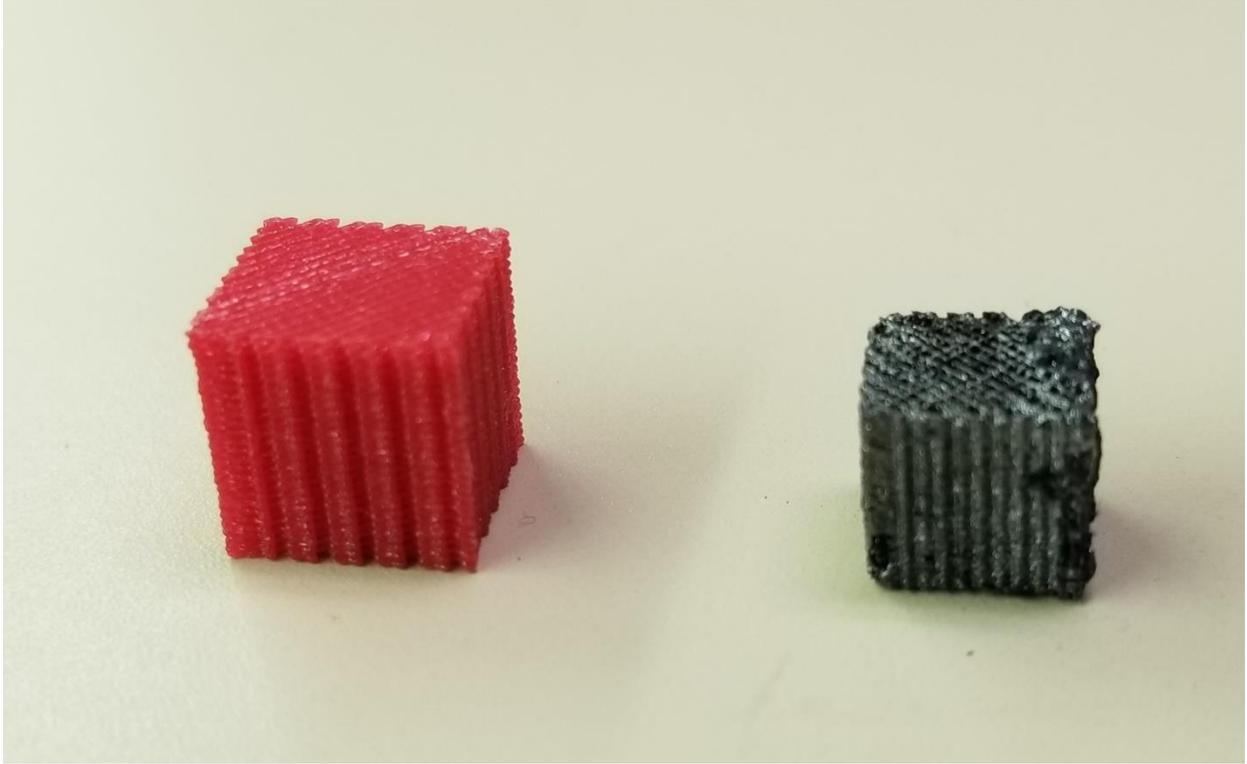

Figure 15: PRO TPE (Matterhackers)

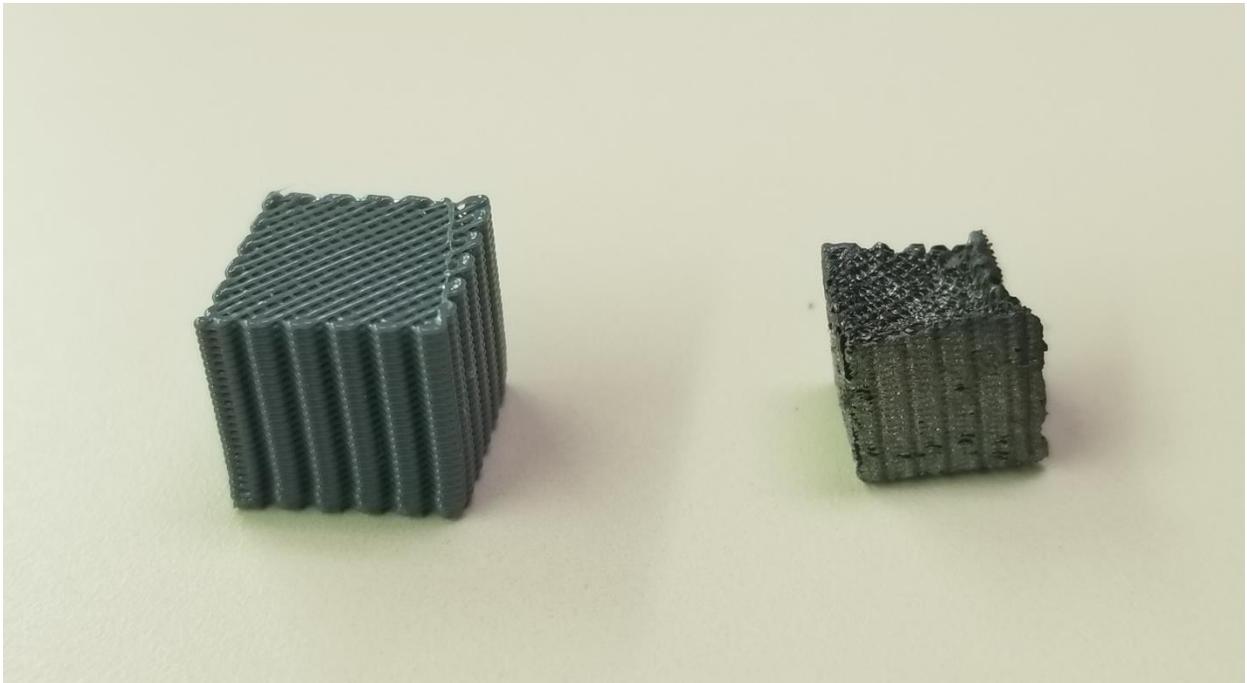

Figure 16: nGen Flex (Colorfabb)

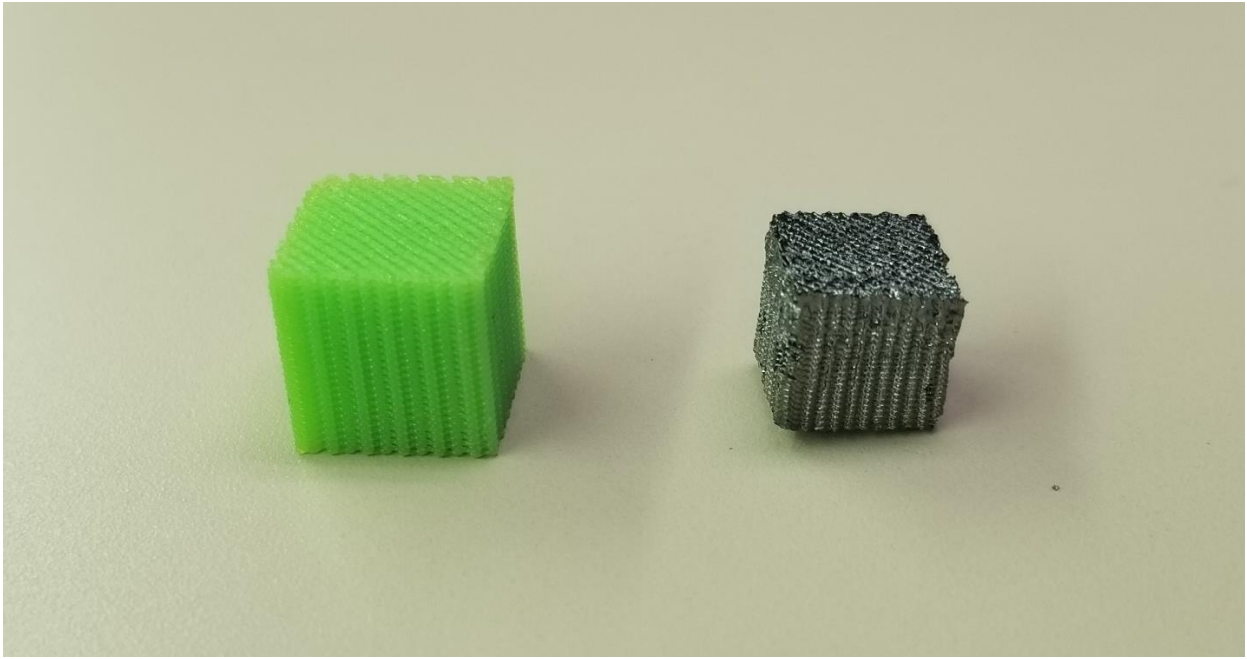

Figure 17: NinjaFlex (NinjaTek)

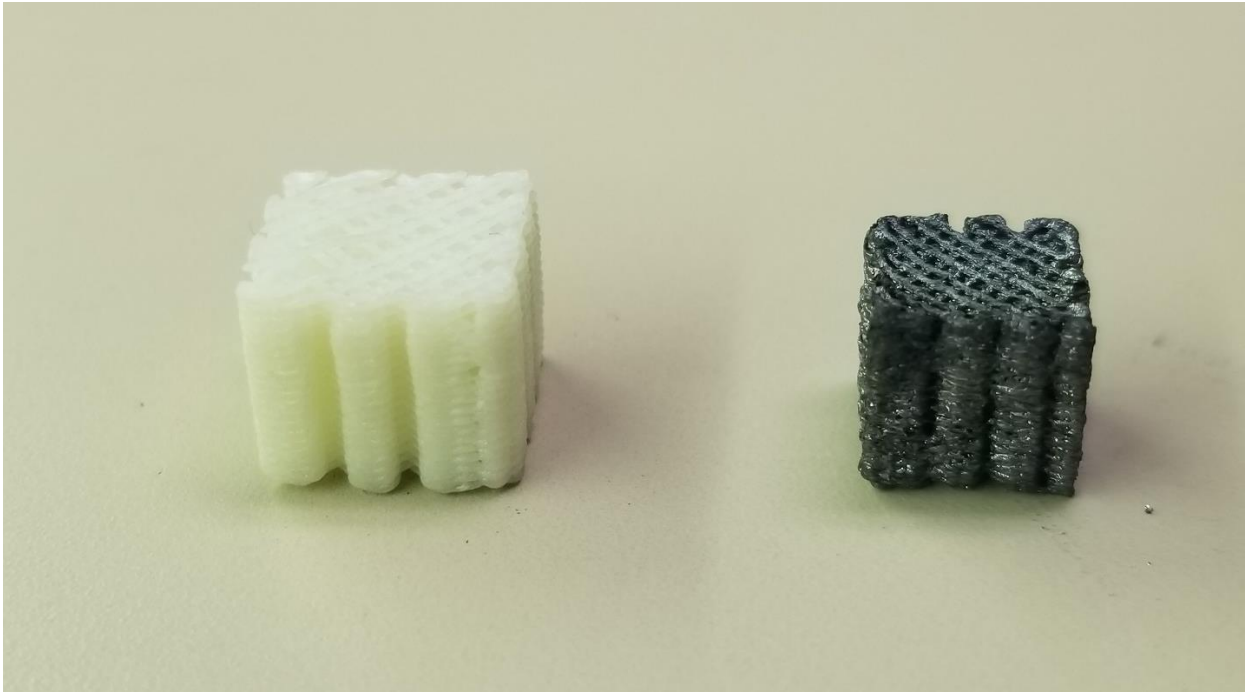

Figure 18: PORO-LAY LAYFOMM 60

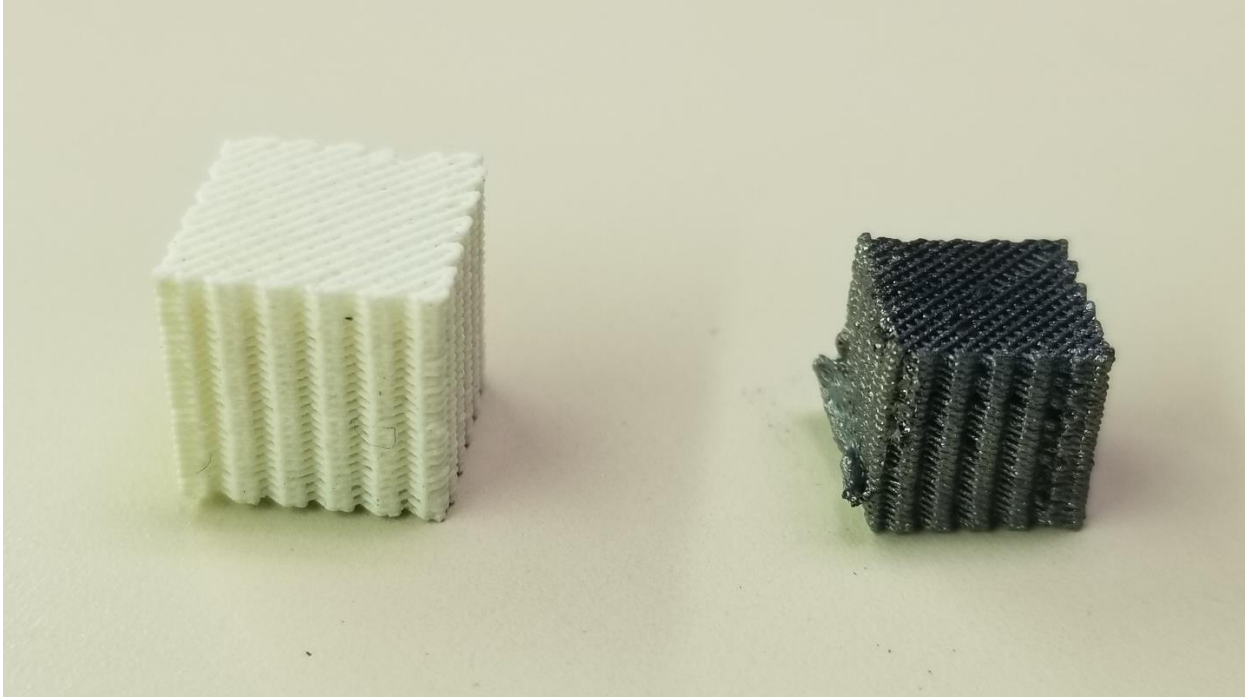

Figure 19: Rubberlay Solay

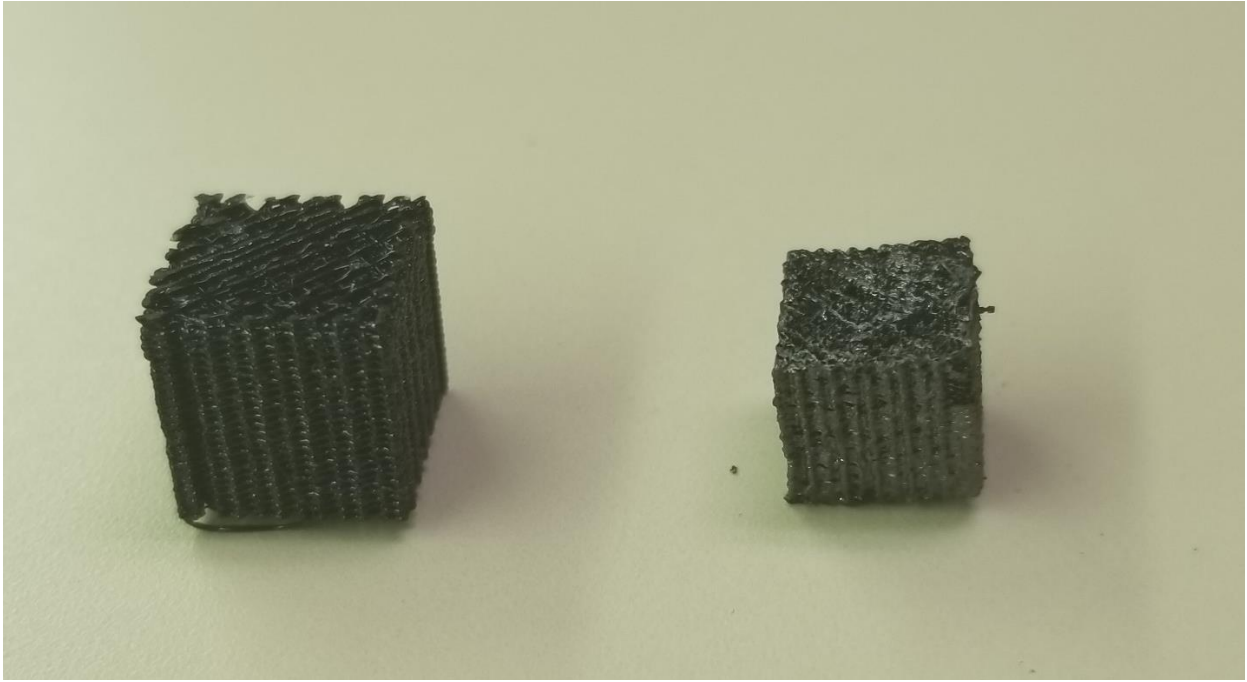

Figure 20: Flexfill 98A ( Fillamentum)

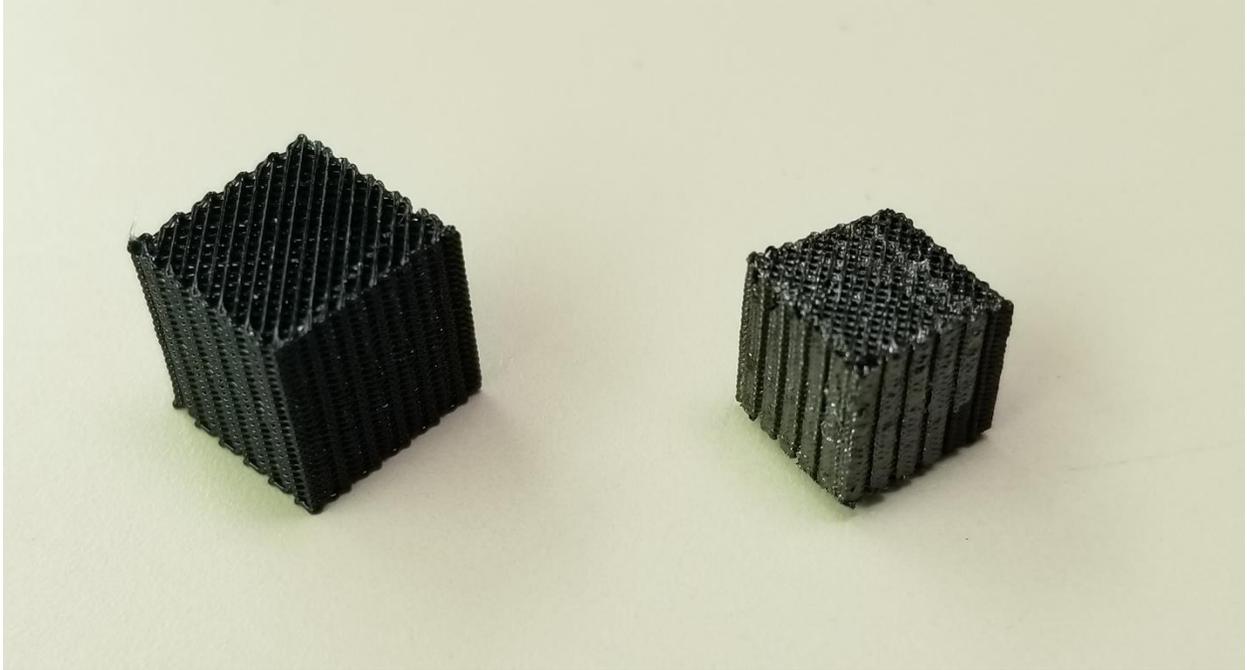

Figure 21: FlexSolid (MadeSolid)

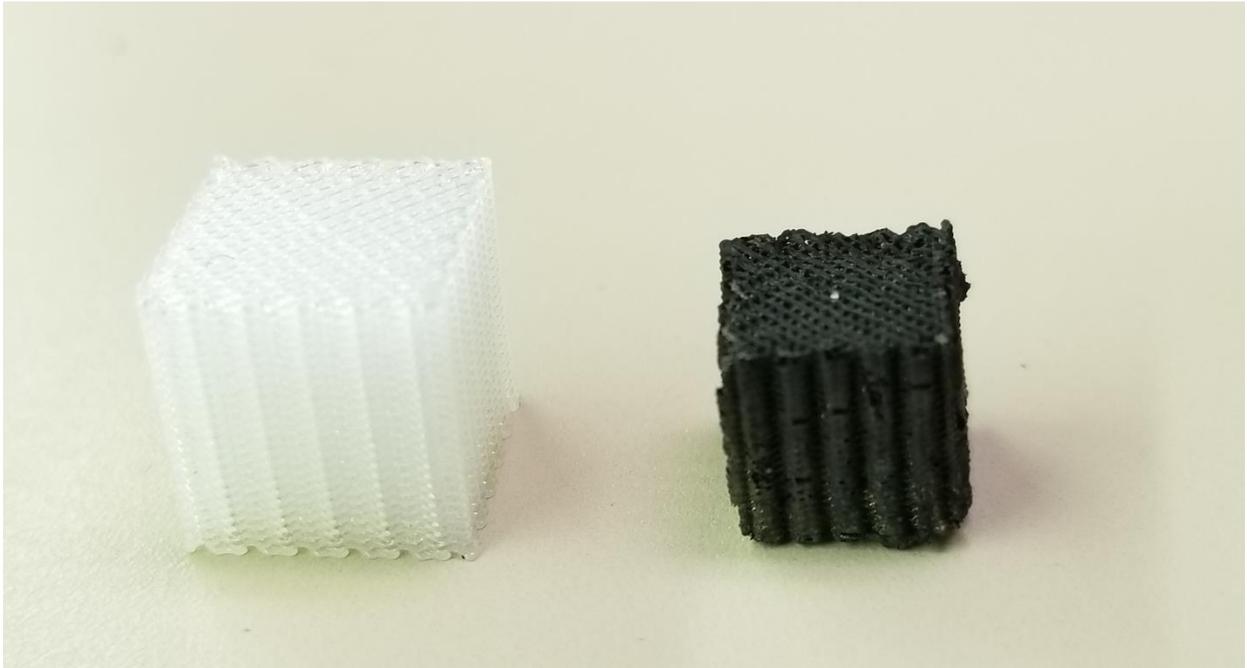

Figure 22: PCTPE (Taulman)

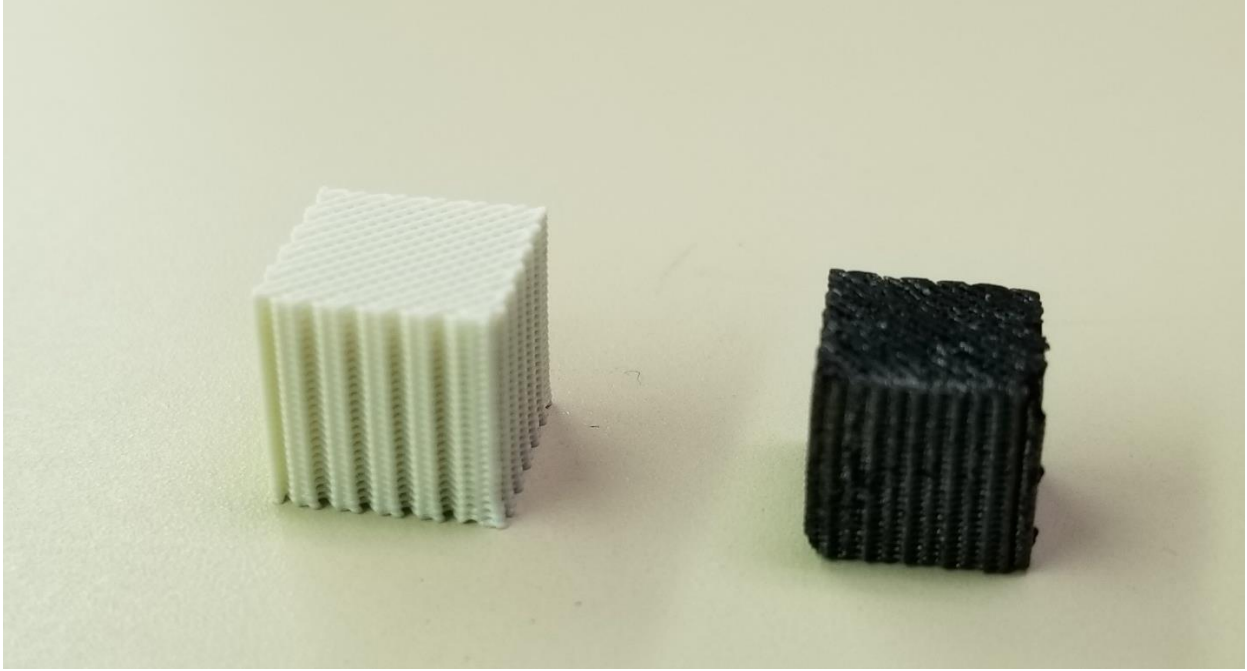

Figure 23: Soft PLA (Matterhackers)

Figure 24 (a) shows the CAD model of the 3-D printed sample and then in (c) the theoretical cross section after pyrolysis with the remaining material representing only the thin surface layer of ceramic present. As can be seen by the cross-sectional SEM images the theoretical prediction was shown to be accurate. Figure 24 (b) shows the SEM of the as-printed PC and then the remaining ceramic is shown after processing in Figure 24 (d). It can be clearly seen that the space initially occupied by polymer filaments is completely empty and ceramic coating is being applied on the surface of the struts creating a thin ceramic structure with far more surface area than the parent 3-D printed structure.

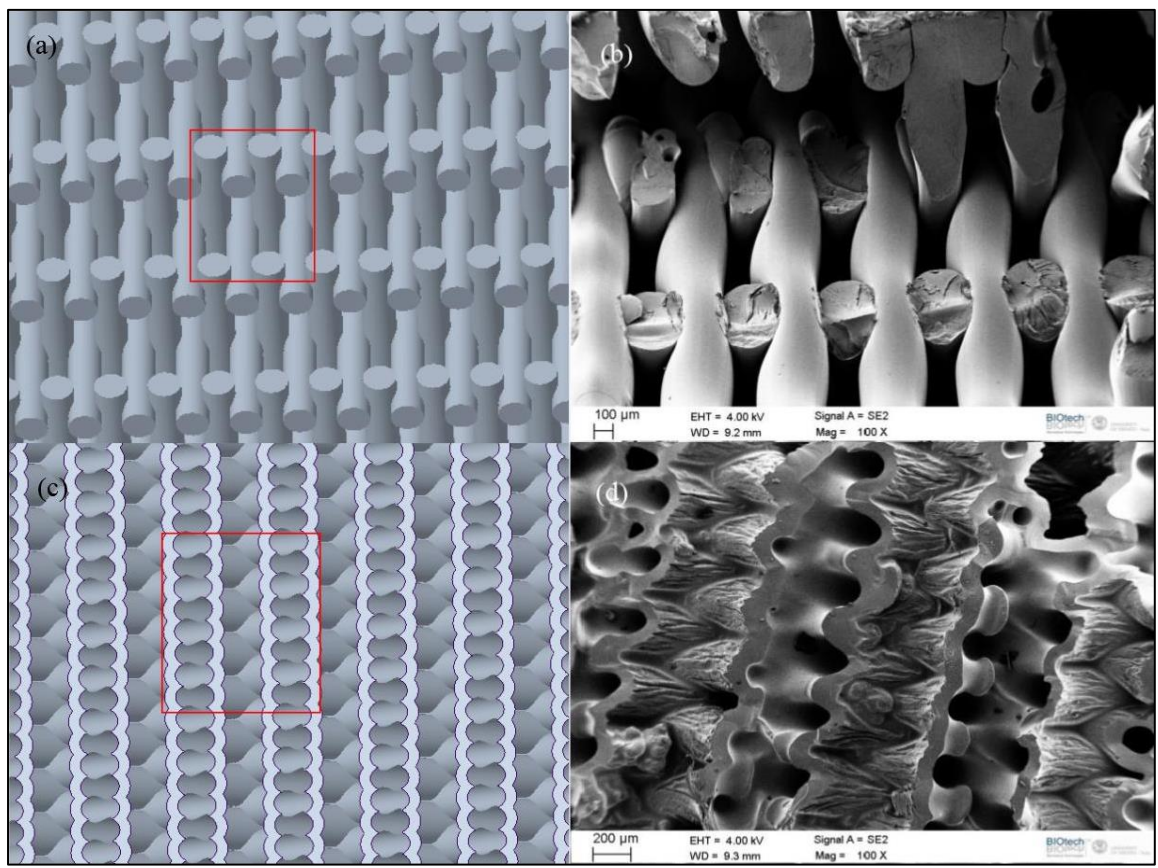

Figure 24: (a) Section of a CAD model of the 3-D printed sample (b) SEM image of a section of polycarbonate (eSUN) sample (c) Section of a CAD model after pyrolysis (d) SEM image of a section of ceramic sample obtained from polycarbonate.

Different filaments/ polymers were observed to produce different results in terms of weight loss, which directly relates to the thickness of the ceramic coating. The ability of the 3-D printed filament to retain the preceramic polymer on its surface can be affected by different parameters such as chemistry of the 3-D printing polymers and the preceramic polymer and also the surface texture of the 3-D printed polymers. It can be seen that Rubberlay (Figure 25(a)) having a very rough surface, produced a thicker coating of the ceramic. Even though Soft PLA (Figure 25(c)) is not as rough as the Rubberlay filament it produced thicker coating than relatively smooth filaments like PP and PETG (shown in Figure 26) which were observed to have a very thin coating on the surface after pyrolysis.

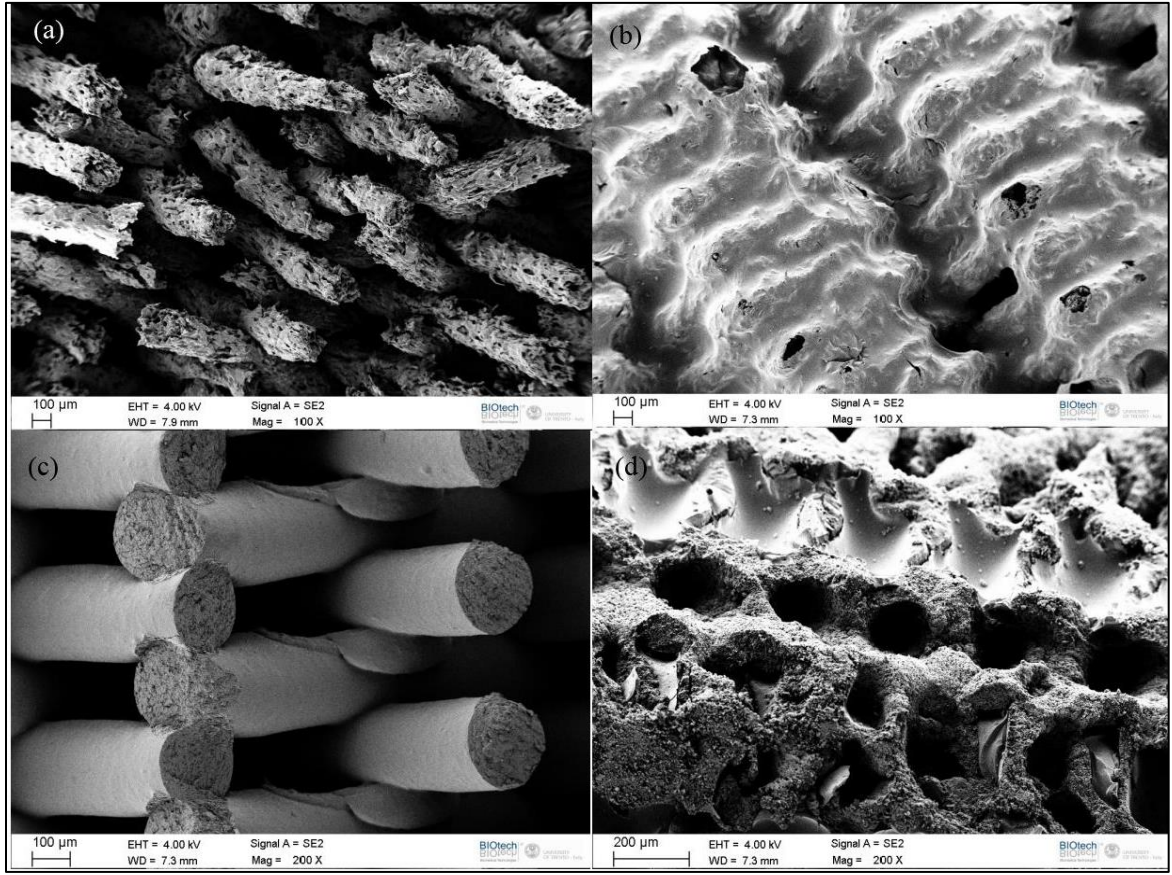

Figure 25: (a) SEM image of a section of the sample printed with Rubberlay filament (b) SEM image of section of ceramic sample obtained from Rubberlay filament (c) SEM image of a section of the sample printed Soft PLA filament (d) SEM image of section of ceramic sample obtained from Soft PLA filament

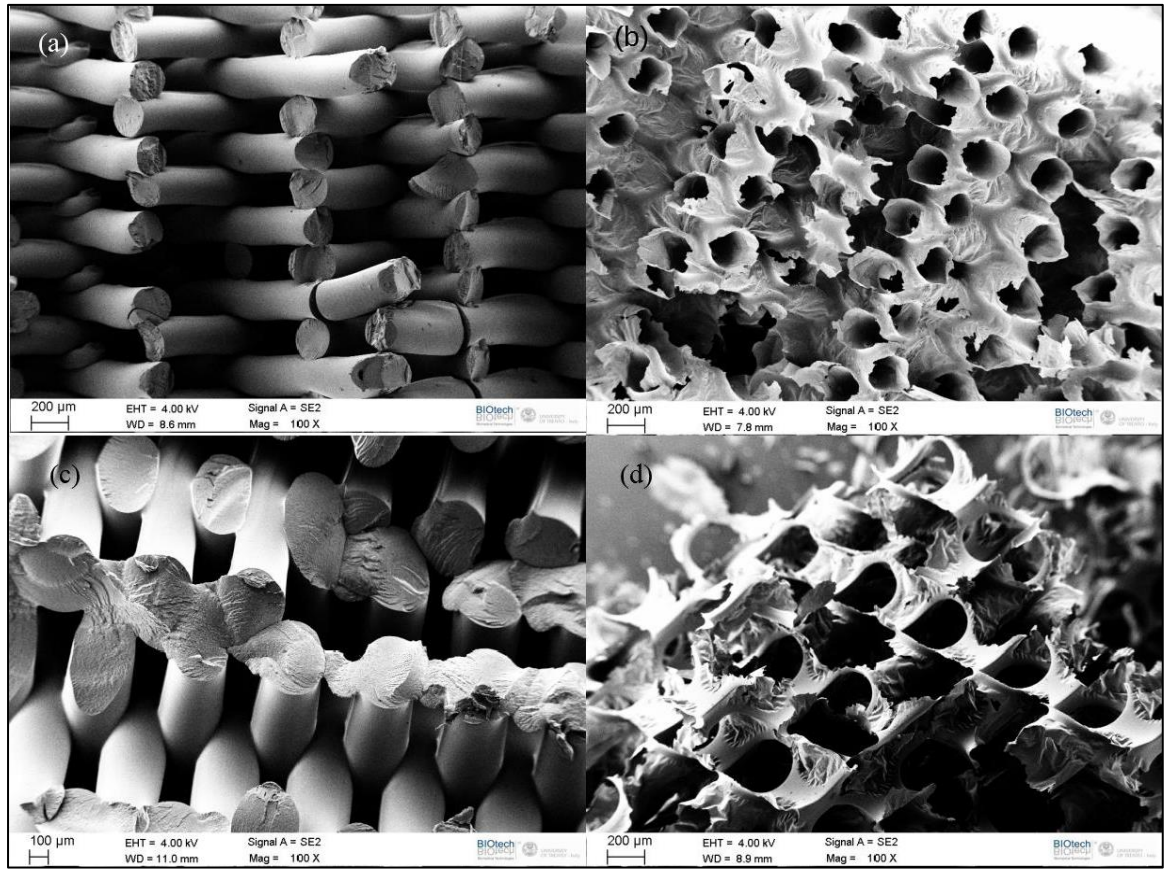

Figure 26: (a) SEM image of a section of the sample printed with PETG filament (b) SEM image of section of ceramic sample obtained from PETG filament (c) SEM image of a section of the sample printed PP filament (d) SEM image of section of ceramic sample obtained from PP filament.

## 4. Discussion

 The results of this experimental study open up a completely new avenue in low-cost 3-D printing of ceramic structures with FFF based methods. The cost of materials used for FFF 3-D printing are much less compared to the cost of powder feedstock required for the laser sintering of the ceramics or the resin slurry required for stereolithography [42]. The equipment itself used for FFF based 3-D printing is much less expensive than any of the other methods with reliable DIY kits costing a few hundred dollars [43] and the Lulzbot FFF printers used in this study costing $2500. This makes the ability to AM in ceramics accessible for a much wider range of individuals, schools, fab labs, makerspaces, community centers, libraries, and small and medium sized enterprises.

As the studied preceramic polymer only coats the outer surface of the 3-D printed polymers, the output ceramic structure consists of two porous systems. One being the designed porosity in the cellular structure and other being the porosity created by decomposition of the polymer filaments during pyrolysis. This type of method is being employed for the first time to create a multi-level porous system using FFF 3-D printing. The ability of the method to apply thin coat of ceramic on to a 3-D printed structures, provides ceramics with high surface to volume ration having high geometric stability.

The ceramic structures produced in this study (silicon oxycarbides) can withstand very high temperatures up to 1400°C [44]. However, since this process is not restricted to the use of polysiloxane - precursor for SiOCs. If polysizanes or polycarbosilane are employed, SiCN and SiC ceramics would be obtained, which are stable up to even higher temperature, in case of SiC up to 1900°C.

By choosing from 3-D printing polymers with different potential to retain the preceramic polymer solution provides (as shown in this screening study), an optimum and consistent thickness of ceramic struts can be obtained for the desired application. If thick struts with more structural strength are needed, polymers like Rubberlay and SoftPLA can be used. Any application where the strength of the ceramic is secondary and very thin ceramic layers are needed, polymers such as PETG and PP can be selected. This selection is based on the processing parameters used, clearly future work could increase or decrease the thickness of the ceramic by adjusting the process parameters.

The components produced in this study can find further applications like ceramic heat exchangers where more surface area is required for heat dissipation [45]. The high resolution of the structures can also enable the use of the study in applications like chemical or gas filters such as catalytic converters [46-48]. The structures can be used in biomedical applications as scaffoldings for bone tissue growth as ceramics have been proven to be compatible as implant materials [49,50].

This study did not attempt to fabricate monolithic ceramic parts with the help of 3-D printed PDCs, since the soaking recipes investigated here will only coat the outer surface and will not impregnate the solid 3-D printed part. Future work is needed to look in more detail to choose the optimum polymer/pre-ceramic polymer system in order to maximize the swelling of the 3-D printed struts and, together with optimum heating, soak times and catalyst concentrations, achieve a dense solid part.

The shrinkage during the conversion from 3-D printed polymer structure to the final ceramic structure needs to be quantified to determine the impact on the geometric integrity of different types of ceramic parts. In addition, the resolution of the FFF based 3-D printing process can be limiting in certain applications. In this study, the smallest head used was 0.25 mm. However, with only the surface PDC method results of thickness less than 100 microns were obtained on the walls as seen in Figure 26 (d).

Future work of the study includes producing finer structures by increasing the 3-D printing resolution by using 3-D printing nozzles smaller than 250 microns. In addition, changing the recipe for the PDCs to make the surface coating even thinner. In order to decrease the cost of the process crosslinking with photoinitiator and curing it with UV light instead of the platinum catalyst should be explored as well as direct printing with a hybrid uv-assisted extrusion process [51]. In the other direction, different preceramic polymers such as polysilazanes or polysilanes, which might be more compatible with the 3-D printing polymers can be used to try and produce 100% dense ceramic cellular structures.

## 5. Conclusions

In summary, the results of this study show that low-cost FFF-based 3-D printing can fabricate 3-D ceramic structures with line widths of 200-250 microns using the PDC process. This also enables accessibility to produce custom ceramics fabricated with geometric stability and high temperature and strength. All of the 3-D printing polymers (PLA, PC, nylon, PP, PETG, PET, copolyesters, TPE, flexible PLA, and TPU) produced definite ceramic structures with varying weight and volume reductions. The filaments with the ability to retain the preceramic such as SoftPLA and Rubberlay produced better results whereas filaments such as PP and PETG were not able to retain the preceramic polymer after soaking as well compared to other filaments. The low-cost processes outlined in this study enable hollow ceramic skins to be fabricated (from all 3-D printing FFF materials) successfully producing ceramics skins of less than100 microns in thickness. The novel results developed here can be used to choose FFF-based polymers to use for PDC processing on a wide range of applications such as heat exchangers, heat sinks, scaffoldings for bone tissue growth, chemical/ gas filters and custom scientific hardware.

### Acknowledgements

This work was partially supported by the funding from University of Trento, the Witte Endowment and Aleph Objects. The authors thank Darren Welson for helpful discussions and technical assistance.